\documentclass[preprint,showpacs,preprintnumbers,amsmath,amssymb,nofootinbib]{revtex4}

\usepackage{dcolumn}
\usepackage{bm}
\usepackage{braket}
\usepackage{amsmath}
\usepackage{txfonts}
\usepackage[pdftex]{graphicx,color}
\usepackage{float}

\begin{document}
\preprint{KUNS2533}
\preprint{KOBE-TH-14-14}

\title{Chromo-Natural Inflation in the Axiverse}

\author{Ippei Obata$^{1,2}$}
\email{obata@tap.scphys.kyoto-u.ac.jp}
\author{Takashi Miura$^{2}$}
 \email{takashi.miura@people.kobe-u.ac.jp}
\author{Jiro Soda$^{2}$}
 \email{jiro@phys.sci.kobe-u.ac.jp}

\affiliation{
$^{1}$Department of Physics, Kyoto University, Kyoto, 606-8502, Japan\\
$^{2}$Department of Physics, Kobe University, Kobe, 657-8501, Japan
}

\collaboration{CLEO Collaboration}

\date{\today}

\begin{abstract}
 We study chromo-natural inflation in the axiverse.
More precisely, we investigate natural inflation with two axions coupled with a SU(2) gauge field.
 Assuming a hierarchy between the coupling constants,
 we find that for certain initial conditions, conventional natural inflation commences and continues for tens of e-foldings, and subsequently 
 chromo-natural inflation takes over from natural inflation.
 For these solutions,  we expect that the predictions are in agreement with observations on CMB scales.
 Moreover, since chromo-natural inflation occurs in the latter part of the inflationary stage, 
chiral primordial gravitational waves are produced 
 in the interesting frequency range higher than $10^{-11}$Hz, which might be detectable by future gravitational wave observations.
\end{abstract}

\pacs{Valid PACS appear here}

\maketitle

\section{Introduction}

 As is well known, an inflationary scenario \cite{Guth:1980zm} resolves issues of the standard big-bang model such as the horizon problem 
and, more importantly, gives rise to explanation of the origin of the anisotropy of the cosmic microwave background radiation (CMB) and large scale structures of the universe.
 However, there exists no conclusive model for inflation based on particle physics.
 The difficulty stems from the fact that we need fine tunings of the inflaton potential to realize slow roll inflation  
 and reconcile resultant predictions with observations.
 In fact, from particle physics point of view, it is difficult to keep these fine tunings against radiative corrections.
 Especially, the mass of a scalar field is quite sensitive to the quantum loop corrections.
 Thus, we need some symmetry to protect the potential from radiative corrections.

 Natural inflation~\cite{Freese:1990rb} is proposed as a solution of the fine-tuning problem. 
 There, the shift symmetry of the axion protect the mass from radiative corrections.
 This symmetry slightly breaks down due to the non-perturbative quantum effect and consequently an appropriate  periodic potential is generated \cite{Svrcek:2006yi}.
 Provided a super-Planckian axion decay constant, it turned out that natural inflation can explain the CMB observations such as the spectral index~\cite{Ade:2013zuv}. 
 However, it would be difficult to realize the super-Planckian axion decay constant in the context of superstring theory or any other fundamental theory~\cite{Freese:2004un, Banks:2003sx}.
 To resolve the issue, we need models where the effective axion decay constant is super-Planckian although the actual axion decay constant is sub-Planckian.
 So far, all of the models proposed to achieve this aim resorted to multi-field generalization of natural inflation. 
 For example,  aligned inflation~\cite{Kim:2004rp}, monodromy inflation~\cite{Silverstein:2008sg}, and N-flation~\cite{Dimopoulos:2005ac} (see also a related work \cite{Jokinen:2004bp}) can be categorized into natural inflation with multiple scalar fields.
Extra natural inflation utilize a component of gauge fields~\cite{ArkaniHamed:2003wu}.
 Moreover, chromo-natural inflation \cite{Adshead:2012kp} is natural inflation with a SU(2) gauge field
where slow-roll inflation is realized by the coupling between the axion and the gauge field.
This is also a kind of multi-field extension of natural inflation.
 Interestingly, apparently different gauge-flation model and non-canonical single field inflation model also belong to this class~\cite{Adshead:2012qe, SheikhJabbari:2012qf,Maleknejad:2011jw, Dimastrogiovanni:2012st} (see also a review \cite{Maleknejad:2012fw}).
 Similar idea is also proposed for ablelian models~\cite{Anber:2009ua}.

Remarkably, chromo-natural inflation has a peculiar feature that sizable chiral gravitational waves can be produced during inflation.
This happens because of the transient tachyonic instability due to the CP violating axion coupling to the gauge field.
Note that the instability occurs only for one of the helicity modes because of CP violation. 
Note that this phenomenon appears in inflation models which have an axion coupling not only with non-abelian but also with abelian gauge fields~\cite{Lue:1998mq, Sorbo:2011rz, Dimastrogiovanni:2012ew, Adshead:2013qp, Adshead:2013nka, Namba:2013kia, Mukohyama:2014gba}\footnote{The conclusion in \cite{Mukohyama:2014gba} that sizable primordial gravitational waves could be produced on CMB scales without any conflict with observations is recently challenged by \cite{Ferreira:2014zia}.}.
 However, in the framework of chromo-natural inflation, either too much gravitational waves are produced or 
 large non-Gaussianity of curvature perturbations are created during inflation.
Indeed, there is no phenomenologically allowed region in the parameter space~\cite{Dimastrogiovanni:2012ew,Adshead:2013qp, Adshead:2013nka, Namba:2013kia}.
 Hence, it is legitimate to say chromo-natural inflation with a single axion is not viable from phenomenological point of view.

 In the spirit of multi-field extension, however, it is natural to consider chromo-natural inflation with multiple axions.
 Indeed,  there appear many axions with various decay constants and coupling constants in view of the superstring theory landscape,
which is dubbed  axiverse~\cite{Arvanitaki:2009fg}.
 Recently, multi-natural inflation has been intensively studied~\cite{Kim:2004rp, Kappl:2014lra, Dimopoulos:2005ac}.
 Nevertheless, the effect of gauge fields is ignored in the setup of the multi-natural inflation.
 In this paper, we study chromo-natural inflation with two axions
  and investigate if sizable chiral gravitational waves can be produced without conflicting with CMB observations.
 In fact, we find that conventional natural inflation occurs on CMB scales, which enables us to circumvent large non-gaussianity 
 and overproduction of gravitational waves on these scales.
 Moreover, we show that chromo-natural inflation commences after tens of e-foldings, which implies that
  chiral primordial gravitational waves become sizable in the interesting frequency range where pulsar timing observation and interferometer
 detectors are available for detection of them.

 This paper is organized as follows: In section 2, we present a chromo-natural inflation model with two axion fields
 and derive equations of motions for the homogeneous fields. 
 In section 3, we investigate the inflationary dynamics. We find that natural inflation occurs first and chromo-natural inflation takes over
 from natural inflation for a class of initial conditions.
 In section 4, we evaluate chiral primordial gravitational waves produced during inflation.
 It turns out that the amplitude of chiral gravitational waves is enhanced on small scales, which might be detectable by future observations.
 The final section is devoted to conclusion.  

\section{Chromo-Natural Inflation with Two Axions}

 In this section, we present an inflationary model in the axiverse and derive equations of background motions.
 Specifically, we consider two axionic fields $\chi$ and $\omega$ which couple with a SU(2) gauge field $A_\mu^a$.
 The field strength of the gauge field $F^a_{\mu\nu}$ is defined by
\begin{equation}
F^a_{\mu\nu} = \partial_\mu A^a_\nu - \partial_\nu A^a_\mu + g\epsilon^{abc}A^b_\mu A^c_\nu \ ,
\end{equation}
where $g$ is its gauge coupling constant and $\epsilon^{abc}$ is the Levi-Civita symbol whose components are structure constants of SU(2) gauge field.
The action reads
\begin{equation}
S=\int dx^4\sqrt{-g}\left[\dfrac{1}{2}R-\dfrac{1}{2}(\partial_\mu\chi)^2-\dfrac{1}{2}(\partial_\mu\omega)^2 -V(\chi,\omega) -\dfrac{1}{4}F^{a\mu\nu}F^a_{\mu\nu}-\dfrac{1}{4}\left(\lambda_\chi\dfrac{\chi}{f}+\lambda_\omega\dfrac{\omega}{h}\right)\tilde{F}^{a\mu\nu}F^a_{\mu\nu} \right] \ ,
\end{equation}
where we used unites $\hbar=c=1$ and ~$M_{pl}=(8\pi G)^{-1/2}=1$ .
 Here, $g$ is a determinant of a metric $g_{\mu\nu}$ (note that it is not related with the gauge coupling constant), $R$ is a Ricci scalar,
 and  $(f, h)$ are decay constants of axions.
 The dual field strength tensor $\tilde{F}^{a\mu\nu}$ is defined by 
\begin{equation}
\tilde{F}^{a\mu\nu} = \dfrac{1}{2!}\epsilon^{\mu\nu\rho\sigma}F^a_{\rho\sigma} \ , \qquad \epsilon^{0123} = \dfrac{1}{\sqrt{-g}} \ .
\end{equation}
 We introduced coupling constants of axions to the gauge field ~$\lambda_\chi \ , \ \lambda_\omega$ .
 Here, we assume that there exists a hierarchy between ~$\lambda_\chi$ ~and ~$\lambda_\omega$ value : $\lambda_\chi \sim {\cal O}(1) \ll \lambda_\omega$ .
 Then we can practically set 
\begin{equation}
\lambda_\chi = 0 \ .
\end{equation}
 The potential for axions $V(\chi,\omega)$ is given by
\begin{align}
V(\chi,\omega) &\equiv \mu_1^4\left[1-\cos{\left(\dfrac{\chi}{f}\right)}\right]+\mu_2^4\left[1-\cos{\left(\dfrac{\omega}{h}\right)}\right] \notag \\
&\equiv U(\chi) + W(\omega) \ , \label{eq: potential}
\end{align}
where $\mu_1, \ \mu_2$ are dynamically generated energy scales.
 We assume that the energy scales of two axions are the same:
\begin{equation}
\mu_1 = \mu_2 \equiv \mu \ .
\end{equation}
We verified that there is no qualitative difference even in the presence of the difference in energy scales 
as long as its hierarchy is not so large. 

Let us consider homogeneous background dynamics in this set-up.
As to the metric, we use a spatially flat metric
\begin{equation}
ds^2=-N(t)^2dt^2+a(t)^2\delta_{ij}dx^idx^j,
\end{equation}
where $N$ is the lapse function.
 After taking the variation of the action, we set $N=1$ in order to regard the time function $t$ as the proper time of hyper surface.
 The axions are homogeneous $\chi = \chi(t), \ \omega = \omega(t)$.
 As a gauge condition, we choose the temporal gauge
\begin{equation}
A^a_0 = 0 \ .
\end{equation}
We also take an ansatz
\begin{equation}
A^a_i=a(t)\phi(t)\delta^a_i \ ,
\end{equation}
which is invariant under the diagonal transformation of the spatial rotation SO(3) and the SU(2) gauge symmetry. It is known that this configuration is dynamically stable~\cite{Maleknejad:2011jr}.
 Thus, their field strength ~$F^a_{\mu\nu}$ ~can be deduced as
\begin{equation}
F^a_{0i}=\dfrac{d(a\phi)}{dt}\delta^a_i \ , \qquad F^a_{ij}=g\epsilon^{abc}A^b_iA^c_j=ga^2\phi^2\epsilon^a_{ij} \ .
\end{equation}
 Substituting these configurations into the action, we obtain the following background action
\begin{equation}
S=\int d^4x\dfrac{a^3}{N}\left[ -3\dfrac{\dot{a}^2}{a^2}+\dfrac{1}{2}\dot{\chi}^2+\dfrac{1}{2}\dot{\omega}^2 - N^2V +\dfrac{3}{2}\dfrac{\dot{(a\phi)}^2}{a^2}-\dfrac{3}{2}N^2g^2\phi^4 - 3Ng\dfrac{\lambda_\omega}{h}\omega\dfrac{\phi^2}{a}\dot{(a\phi)} \right] \ ,
\end{equation}
where a dot denotes a derivative with respect to the cosmic time $t$ .
 Taking the variation with respect to ~$N$ and setting $N=1$ after the variation,  we obtain the Hamiltonian constraint
\begin{equation}
3H^2=\dfrac{1}{2}\dot{\chi}^2+\dfrac{1}{2}\dot{\omega}^2+\dfrac{3}{2} \dot{(a\phi)}^2 a^{-2}+\dfrac{3}{2}g^2\phi^4+V \ ,
\end{equation}
where ~$H \equiv \dot{a}/a$ ~is the Hubble parameter.
 The equations for inflatons and gauge fields read
\begin{gather}
\ddot{\chi} + 3H\dot{\chi} + U_{\chi} = 0 \ , \label{eq: chimotion}\\
\ddot{\omega} + 3H\dot{\omega} + W_{\omega} = - 3\dfrac{\lambda_\omega}{h}g\phi^2(\dot{\phi} + H\phi) \ , \label{eq: omegamotion}\\
\ddot{\phi}+3H\dot{\phi}+(\dot{H}+2H^2)\phi+2g^2\phi^3 = g\dfrac{\lambda_\omega}{h}\phi^2\dot{\omega} \label{eq: gaugemotion}\ .
\end{gather}
 Here, we defined
\begin{equation}
U_\chi \equiv \dfrac{dU}{d\chi} = \dfrac{\mu^4}{f}\sin{(\dfrac{\chi}{f})} \ , \qquad W_\omega \equiv \dfrac{dW}{d\omega} = \dfrac{\mu^4}{h}\sin{(\dfrac{\omega}{h})} \ .
\end{equation}
 The equation for the scale factor ~$a(t)$ ~ is given by
\begin{equation}
\dot{H} = - \dfrac{1}{2}\dot{\chi}^2 - \dfrac{1}{2}\dot{\omega}^2 - (\dot{\phi} + H\phi)^2 - g^2\phi^4 \ .
\end{equation}
 We can expect that ~$\chi$ ~plays a role of inflaton for natural inflation, ~$\omega$ ~becomes an inflaton for chromo-natural inflation.
 We will see that both types of inflation can occur for appropriate initial conditions.

\section{An inflationary dynamics}

 In this section, we discuss an inflationary trajectory of this two-field inflation. We
 perform numerical calculations  with the following sets of parameters: 
$( \ f, \ h, \ \mu, \ g, \ \lambda_\omega \ ) = ( \ 5, \ 5\times10^{-4}, 10^{-2}, \ 10^{-3}, \ 1.5\times10^{3} \ )$.
 Note that this example is not the only way to realize our set-up.

 We recall the property of each inflation at first.
 From \eqref{eq: chimotion} the slow-roll equation for ~$\chi$ ~reads
\begin{equation}
3H\dot{\chi} + U_\chi \approx 0 \label{eq: slowchi} \ .
\end{equation}
 This is a conventional single field slow-roll equation.
 If natural inflation is dominant,  slow-roll parameters in terms of  the potential are defined  by
\begin{align}
\epsilon_V \approx \dfrac{1}{2}\left(\dfrac{U_\chi}{U}\right)^2 \ , \quad \eta_V \approx \dfrac{U_{\chi\chi}}{U} \ ,
\end{align}
where ~$U_{\chi\chi}$ ~is the second order derivative of ~$U(\chi)$ ~with respect to ~$\chi$ .
 On the other hand, from Eqs. \eqref{eq: omegamotion} and \eqref{eq: gaugemotion},
 slow-roll equations for $\omega$ and the gauge field read
\begin{gather}
3H\dot{\omega} + W_{\omega} \approx - 3\dfrac{\lambda_\omega}{h}g\phi^2(\dot{\phi} + H\phi) \ , \label{eq: omegamotion2}\\
3H\dot{\phi} + 2H^2 \phi + 2g^2\phi^3 \approx g\dfrac{\lambda_\omega}{h}\phi^2\dot{\omega} \label{eq: gaugemotion2} \ .
\end{gather}
 Chromo-natural inflation happens when the ``magnetic drift'' term (the coupling term of axion to the gauge sector) is sufficiently large : ~$\lambda_\omega^2 g^2 \phi^4 \gg h^2H^2$ ~(note that ~$g^2\phi^4 \ll H^2$ ~during inflation) .
 Then diagonalising \eqref{eq: omegamotion2} and \eqref{eq: gaugemotion2} for ~$\dot{\omega}$ ~and ~$\dot{\phi}$ ,  we obtain the slow-roll equations
\begin{gather}
\dfrac{\lambda_\omega}{h}\dot{\omega} \approx -\dfrac{hHW_\omega}{\lambda_\omega g^2\phi^4} - \dfrac{H^2}{g\phi} + 2g\phi \label{eq: omegamotion3} \ , \\
\dot{\phi} \approx -H\phi - \dfrac{hW_\omega}{3\lambda_\omega g\phi^2} \ .
\end{gather}
Since ~$\phi$ ~is almost constant during chromo-natural inflation,  its value becomes 
\begin{equation}
\phi \approx \phi_{min} \equiv -\left(\dfrac{hW_\omega}{3g\lambda_\omega H}\right)^{1/3} \label{eq: phimin} \ .
\end{equation}
 Substituting the above value \eqref{eq: phimin} into the first term in the right hand side of Eq. \eqref{eq: omegamotion3},  we 
can deduce the slow-roll equation for  $\omega$  as
\begin{equation}
\dfrac{\lambda_\omega}{h}\dot{\omega} \approx 2g\phi + \dfrac{2H^2}{g\phi} \equiv -2H\dfrac{1 + m_\phi^2}{m_\phi} \label{eq: slowomega} \ ,
\end{equation}
 where we defined the following parameter
\begin{equation}
m_\phi \equiv -\dfrac{g\phi}{H} \ . \label{eq: m_phi}
\end{equation}
 We show that this parameter is relevant to the chiral instability of gauge fluctuations and plays a crucial role for determining the amount of gravitational waves generated by gauge fields.
 From these slow-roll conditions, if chromo-natural inflation is dominant,  slow-roll parameters defined by the Hubble parameter are written by
\begin{align}
\epsilon_H &\approx \dfrac{h}{\lambda_\omega}\dfrac{1 + m_\phi^2}{m_\phi}\dfrac{W_\omega}{W} \ , \\
\eta_H &\approx \dfrac{h}{\lambda_\omega}\dfrac{1 + m_\phi^2}{m_\phi}\left( \dfrac{2W_\omega}{W} - \dfrac{W_{\omega\omega}}{W_{\omega}} \right) \ ,
\end{align}
where ~$W_{\omega\omega}$ ~is the second order derivative of ~$W(\omega)$ ~with respect to ~$\omega$ .

 Then we find that ~$\dot{\omega}$ ~can be very small compared to ~$\dot{\chi}$ ~due to the magnetic drift factor.
 In fact, we can approximate
\begin{equation}
\left |\dfrac{\dot{\omega}}{\dot{\chi}} \right | \approx 2\dfrac{V}{U_\chi}\dfrac{h}{\lambda_\omega }\dfrac{1 + m_\phi^2}{m_\phi} \label{eq: velovelo} \ ,
\end{equation}
where the factor ~$h/\lambda_\omega$ ~must be small in order for chromo-natural inflation to occur. 
 When the rate \eqref{eq: velovelo} is sufficiently small, the inflationary trajectory of this dynamics is almost along $\chi$ direction.
Once the slow-roll conditions for $\chi$  break down, the trajectory goes along $\omega$ direction and chromo-natural inflation occurs.
 From this feature, we can roughly estimate the number of e-foldings  $N_e$  as a sum of the number of e-foldings for each trajectory :
\begin{equation}
N_e\equiv N_{e\chi}+N_{e\omega} \ ,
\end{equation}
where we defined
\begin{align}
N_{e\chi} &\equiv \int_{t_i}^{t_m}Hdt \ , \\
N_{e\omega} &\equiv \int_{t_m}^{t_f}Hdt \approx \int_{\omega_m}^{\omega_f} \dfrac{H}{\dot{\omega}}d\omega \ .
\end{align}
 Note that natural inflation becomes dominant during $t_i < t < t_m$ and chromo-natural inflation occurs until $t = t_f$.
 Unfortunately, it is difficult to evaluate $N_{e\chi}$ analytically because both the dynamics of $\chi$ and $\omega$ are relevant in general.
However, we can estimate $N_{e\omega}$ from the attractor value \eqref{eq: phimin} and the slow roll equation \eqref{eq: slowomega} 
\begin{equation}
N_{e\omega} \approx -\int_{\tilde{\omega}_m}^{\tilde{\omega}_f}\dfrac{3^{2/3}}{2}\dfrac{\mu^{4/3}g^{2/3}\lambda_{\omega}^{4/3}H^{4/3}\sin{\tilde{\omega}}^{1/3}}{3\lambda_{\omega}^{2/3}H^{8/3}+3^{1/3}g^{4/3}\mu^{8/3}\sin{\tilde{\omega}}^{2/3}}d\tilde{\omega} \label{eq:efo} \ ,
\end{equation}
where ~$\tilde{\omega}\equiv\omega/h$ .

Here, we focus on the solutions which realize a natural inflation on CMB scales and a chromo-natural inflation on scales smaller than CMB scales. 
 In general, CMB constraints are determined by the dynamics of fluctuations which cross the horizon at 50-60 e-folds before the end of inflation, $N_{\text{COBE}}$.
 So, if adiabatic scalar fluctuation or its non-Gaussianity are almost derived from natural inflation at $N_{\text{COBE}}$, that is, $\chi$ is dominant contribution of the potential energy at first, we can choose appropriate background values to suppress chiral gravitational waves without overproducing scalar fluctuations \cite{Adshead:2013qp,Adshead:2013nka}.
 In order to realize such a condition, we consider the inflationary trajectory where $N_{e\omega}$ is smaller than $N_{\text{COBE}}$; $N_{e\omega} \lesssim N_{\text{COBE}}$.
 Practically, we can set $\tilde{\omega}_f$ zero. Thus, from \eqref{eq:efo}, we can numerically deduce
 the condition  $\tilde{\omega}_m \lesssim 0.12\pi$
for the parameters $( \ f, \ h, \ \mu, \ g, \ \lambda_\omega \ ) = ( \ 5, \ 5\times10^{-4}, 10^{-2}, \ 10^{-3}, \ 1.5\times10^{3} \ )$.
 Fortunately, for small $\tilde{\omega}$, the number of e-foldings $N_{e\omega}$ can be estimated as  
\begin{align}
N_{e\omega} &\approx \int_{\omega_f}^{\omega_m}\dfrac{\tilde{\omega}}{2}\dfrac{[12\mu^{4}g^{2}\lambda_{\omega}^{4}]^{1/3}}{[\lambda_{\omega}\mu^4\tilde{\omega}^3]^{2/3}+[12g^2]^{2/3}}d\tilde{\omega} \notag \\
                      &= \dfrac{1}{4}\left(\dfrac{12g^{2}\lambda_{\omega}^{2}}{\mu^{4}} \right)^{1/3}\left.\left[\log\left( 1 + \left(\dfrac{\lambda_{\omega}\mu^{4}}{12g^{2}}\right)^{2/3} \tilde{\omega}(t)^2 \right) \right] \right |_{\tilde{\omega}_f}^{\tilde{\omega}_m} \notag \\
                      &= \dfrac{1}{4}\left(\dfrac{12g^{2}\lambda_{\omega}^{2}}{\mu^{4}} \right)^{1/3}\log\left( 1 + \left(\dfrac{\lambda_{\omega}\mu^{4}}{12g^{2}}\right)^{2/3} \tilde{\omega}(t_m)^2 \right) \ .
\end{align}
Substituting the parameters into above formula, we actually see inequality $\tilde{\omega}_m \lesssim 0.12\pi$  holds for
achieving the condition $N_{ew} \lesssim N_{\text{COBE}}$ .

\begin{figure}[h]
\begin{center}
\includegraphics[width=8cm,height=8cm,keepaspectratio]{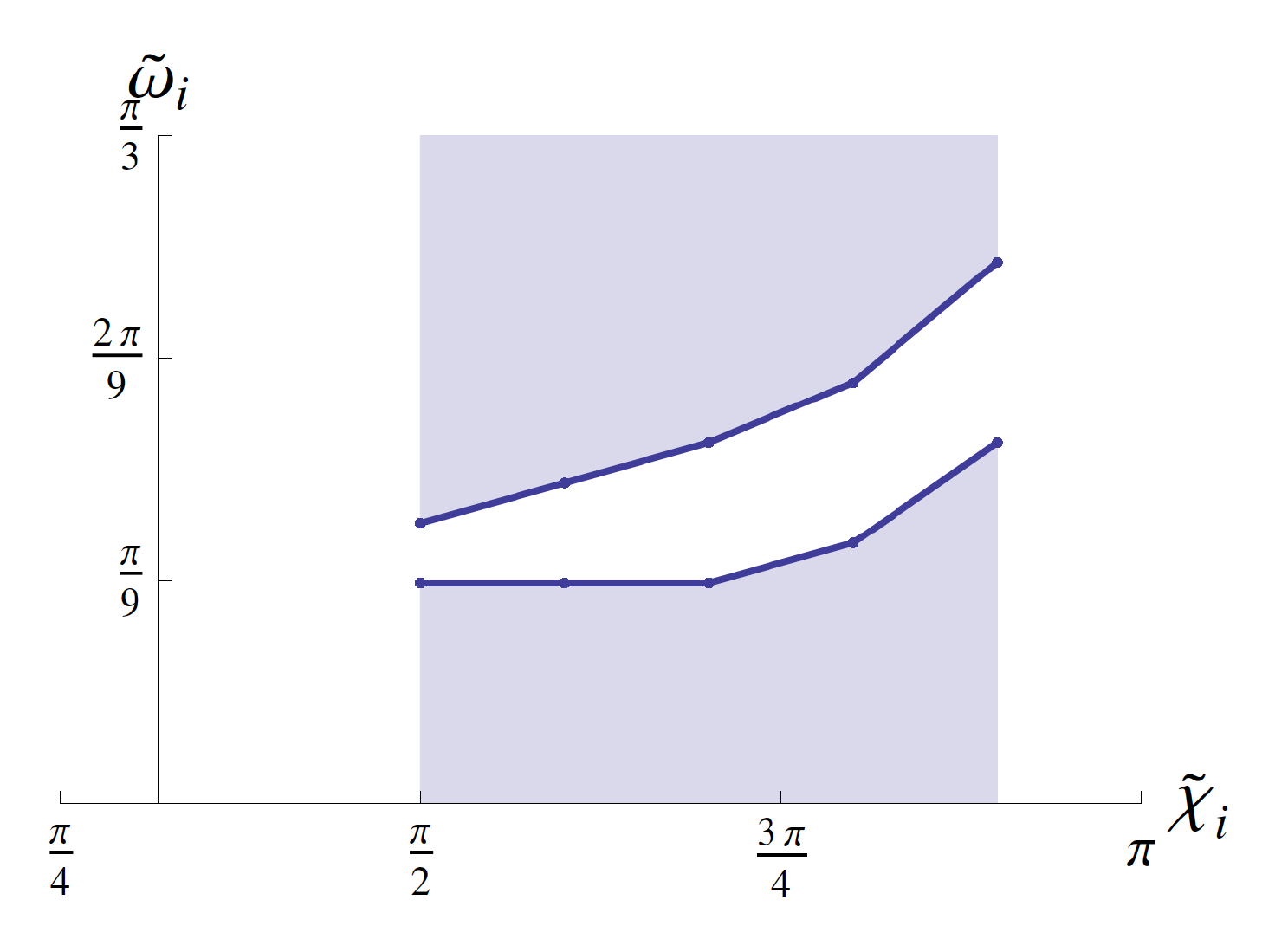}
\end{center}
\caption{The region (white) of initial values which satisfies conditions  $N_e \gtrsim N_{\text{COBE}}$ and $N_{ew} \lesssim N_{\text{COBE}}$ under a restriction
 $\pi/2 \leq \tilde{\chi}_i \leq 9\pi/10 \ (\tilde{\chi}_i \equiv \chi_i/f)$.
 The top blue line represents $N_{ew} \sim N_{\text{COBE}}$ and bottom blue line is the lower bound which satisfies both $N_{ew} \geq 0$ and $N_e \gtrsim N_{\text{COBE}}$.
 Hence in the middle white band we can find phenomenologically viable solutions with chiral gravitational waves.
   }
\label{fig: ini}
\end{figure}

 As to the initial conditions for $\chi$ and $\omega$, we numerically find that there are many cases for which the inequality $N_{ew} \lesssim N_{\text{COBE}}$ holds and natural inflation becomes dominant on CMB scales. In  FIG.\ref{fig: ini}, we plotted the region in the space of initial conditions
where  the conditions $N_e \gtrsim N_{\text{COBE}}$ and $N_{ew} \lesssim N_{\text{COBE}}$ are satisfied.
 For these cases, we expect that chiral gravitational waves on CMB scales are suppressed and all other observational constraints are satisfied.

\section{Chiral Gravitational Waves}

In the previous section, we found chromo-natural inflation can take over from the conventional natural inflation.
Hence, we expect that  sizable chiral gravitational waves are produced in the interesting range of frequencies, which
might be detectable by interferometer detectors  such as advanced LIGO \cite{Abramovici:1992ah} or KAGRA \cite{Somiya:2011np}.
In this section, we perform numerical calculations  with a set of parameters: $( \ f, \ h, \ \mu, \ g, \ \lambda_\omega \ ) = ( \ 3.5, \ 3.5\times10^{-4}, 10^{-2}, \ 10^{-3}, \ 1.5\times10^{3} \ )$.
 Let us examine if this actually occurs or not.

 The metric with tensor perturbations  reads 
\begin{equation}
ds^2 = a(\tau)^2[-d\tau^2 + (\delta_{ij} + h_{ij})dx^idx^j] \ ,
\end{equation}
where ${h^i}_j$ is transverse and traceless, ${h^i}_i={h^{ij}}_{,j}=0$.
 Using a new variable $\psi_{ij}\equiv a(\tau)h_{ij}$, the quadratic action $S_{\text{EH}}$ for the tensor perturbations
  is given by
\begin{equation}
\delta S_{\text{EH}}=\int dx^4\dfrac{1}{2}\left[\dfrac{1}{4}\psi'^{ij}\psi'_{ij}-\dfrac{1}{4}\psi^{ij,k}\psi_{ij,k}-\left(\dfrac{3}{4}\dfrac{a''}{a}-\dfrac{1}{2}\left(\dfrac{a'}{a}\right)^2\right)\psi^{ij}\psi_{ij}\right] \ ,
\end{equation}
where a prime represents a derivative with respect to a conformal time $\tau$.
 From the action for the scalar field $S_{\text{scalar}}$, we also have a contribution to the quadratic action for tensor perturbations
\begin{equation}
\delta S_{\text{scalar}}=\int d^4x \left(-\dfrac{a^2}{4}\psi^{ij}\psi_{ij}\right)\left[ \dfrac{1}{2a^2}(\chi'^2+\omega'^2) - V \right] \ .
\end{equation}
 Next, we define the perturbation for the gauge field as follows:
\begin{equation}
A{^a}_i = a\phi\delta{^a}_i + t{^a}_i \ ,
\end{equation}
where ${t^i}_j$ is also transverse and traceless, ${t^i}_i={t^{ij}}_{,j}=0$.
 Note that we can treat the second term as a tensor since the index "$a$" can be identified with the spatial index "$i$".
 Then, the action $S_{\text{gauge}}$  for the gauge sector is given by
\begin{gather}
\delta S_{\text{gauge}} 
 = \int d^4x\left(-\dfrac{1}{4}\right)\left[-2t'^a_i t'^a_i - \dfrac{1}{2a^2}(a\phi)'^2\psi^{ij}\psi_{ij}+\dfrac{4}{a}(a\phi)'\psi^{ij}t'_{ij}+2t^a_{i,j}t^a_{i,j} - 4ga\phi\epsilon^{abi}t^b_jt^a_{j,i} \right. \notag \\
\left. -4ga\phi^2\psi^{jm}\epsilon^a_{ij}(t^a_{m,i}-t^a_{i,m})-4g^2a^2\phi^3\psi^{ij}t_{ij}+\dfrac{3}{2}g^2a^2\phi^4\psi^{ij}\psi_{ij}+\dfrac{\lambda_{\omega}}{f}\omega \left(2\epsilon^{ijk}t'^a_i(t^a_{k,j}-t^a_{j,k})-2g(a\phi t^{ij}t_{ij})' \right)\right] \ .
\end{gather}
 We can rewrite the total action in terms of Fourier components defined by
\begin{align}
\psi_{ij}(\bm{x},\tau) &= 2\sum_{A = \pm}\int\dfrac{d^3\bm{k}}{(2\pi)^3}e^{A}_{ij}(\bm{k})\psi^A_{\bm{k}}(\tau)e^{i\bm{k}\cdot\bm{x}} \ , \\
t_{ij}(\bm{x},\tau) &= \sum_{A = \pm}\int\dfrac{d^3\bm{k}}{(2\pi)^3}e^{A}_{ij}(\bm{k})t^A_{\bm{k}}(\tau)e^{i\bm{k}\cdot\bm{x}} \ ,
\end{align}
where $e^{A}_{ij}(\bm{k})$ are the polarisation tensors which satisfy the following normalization relation, $e^{Aij}(\bm{k})e^{B}_{ij}(\bm{-k})=\delta^{AB}$, and the index ``$A = \set {+, -}$" represents circular polarisation states defined by $ik^i\epsilon^a_{ij}e^\pm_{jm}(\bm{k})=\pm ke^\pm_{am}(\bm{k})$.
 Thus, we get the following quadratic action for tensor perturbations 
\begin{align}
\delta S_{\text{tensor}} &\equiv \delta S_{\text{EH}}+\delta S_{\text{scalar}}+\delta S_{\text{gauge}} \notag \\
 &= \dfrac{1}{(2\pi)^3}\int d^3\bm{k}d\tau \left[ \dfrac{1}{2}\bar{\psi}^{\pm'}_{\bm{k}} \psi^{\pm'}_{\bm{k}} - \dfrac{1}{2}k^2 \bar{\psi}^\pm_{\bm{k}}\psi^\pm_{\bm{k}} + \dfrac{1}{2} \left( \dfrac{a''}{a} + 2\dfrac{(a\phi)'^2}{a^2} - 2g^2a^2\phi^4 \right) \bar{\psi}^\pm_{\bm{k}}\psi^\pm_{\bm{k}} \right] \notag \\
 &+ \dfrac{1}{(2\pi)^3}\int d^3\bm{k}d\tau \left[  \dfrac{1}{2}\bar{t}^{\pm'}_{\bm{k}} t^{\pm'}_{\bm{k}} - \dfrac{1}{2}k^2 \bar{t}^\pm_{\bm{k}}t^\pm_{\bm{k}} -\dfrac{\lambda_\omega}{2h}ga\phi\omega' \bar{t}^\pm_{\bm{k}}t^\pm_{\bm{k}} \pm \dfrac{1}{2}k\left( 2ga\phi + \dfrac{\lambda_\omega}{h}\omega' \right)\bar{t}^\pm_{\bm{k}}t^\pm_{\bm{k}} \right. \notag \\
 &\left. \mp kga\phi^2(\bar{\psi}^\pm_{\bm{k}}t^\pm_{\bm{k}}+\bar{t}^\pm_{\bm{k}}\psi^\pm_{\bm{k}}) + g^2a^2\phi^3(\bar{\psi}^\pm_{\bm{k}}t^\pm_{\bm{k}}+\bar{t}^\pm_{\bm{k}}\psi^\pm_{\bm{k}}) - \dfrac{(a\phi)'}{a}(\bar{\psi}^\pm_{\bm{k}}t^{\pm'}_{\bm{k}}+\bar{t}^{\pm'}_{\bm{k}}\psi^\pm_{\bm{k}}) \right] \ .
\end{align}
 Then, we get the equations of motion for tensor perturbations
\begin{gather}
\psi^{\pm''}_{\bm{k}} + \left(k^2 - \dfrac{a''}{a} - 2((\phi'+\dfrac{a'}{a}\phi)^2 - g^2a^{2}\phi^4) \right)\psi^\pm_{\bm{k}} = -2\left( ( \pm kga\phi - g^2a^{2}\phi^2 )\phi t^\pm_{\bm{k}} + (\phi'+\dfrac{a'}{a}\phi) t^{\pm'}_{\bm{k}} \right) \ , \label{eq: tenmetric} \\
t^{\pm''}_{\bm{k}}+\left( k^2 + \dfrac{\lambda_\omega ga\phi}{h}\omega' \mp k\left(2ga\phi + \dfrac{\lambda_\omega}{h}\omega'\right)  \right)t^\pm_{\bm{k}}=-2\left( ( \pm kga\phi - g^2a^{2}\phi^2 )\phi\psi^\pm_{\bm{k}} -\left((\phi'+\dfrac{a'}{a}\phi) \psi^\pm_{\bm{k}} \right)' \right) \label{eq: tengauge} \ .
\end{gather}
 Now, we replace $\tau$ with the following dimensionless parameter
\begin{equation}
x\equiv -k\tau \ .
\end{equation}
 Moreover, we can use the following slow-roll conditions 
\begin{gather}
a(\tau) = -\dfrac{1}{H\tau} \ , \\
\phi' = 0 \ , \\
\dfrac{\lambda_\omega}{h}\omega' = -2\dfrac{1 + m_\phi^2}{m_\phi}Ha \ .
\end{gather}
 Then, \eqref{eq: tenmetric} and \eqref{eq: tengauge} can be reduced to
\begin{gather}
\dfrac{d^2\psi^\pm_{\bm{k}}}{dx^2}+\left( 1-\dfrac{2}{x^2} -\dfrac{2}{x^2}(1 - m_\phi^2)\phi^2 \right)\psi^\pm_{\bm{k}} = 2\dfrac{\phi}{x}\dfrac{dt^\pm_{\bm{k}}}{dx} + 2m_\phi (m_\phi \pm x)\dfrac{\phi}{x^2}t^\pm_{\bm{k}} \label{eq: metrictensor} \ , \\
\dfrac{d^2 t^\pm_{\bm{k}}}{dx^2} + \left(1+\dfrac{m}{x^2} \pm \dfrac{m_t}{x} \right)t^\pm_{\bm{k}} = -2\phi\dfrac{d}{dx}\left( \dfrac{\psi^\pm_{\bm{k}}}{x} \right) + 2m_\phi (m_\phi \pm x)\dfrac{\phi}{x^2}\psi^\pm_{\bm{k}} \label{eq: gaugetensor} \ ,
\end{gather}
where we defined
\begin{gather}
m \equiv 2(1+m_\phi^2) \ ,\\
m_t \equiv 2\left( 2m_\phi + \dfrac{1}{m_\phi} \right) \ .
\end{gather}

 From these equations, we notice that the dynamics of tensor perturbations depends on the helicity.
 For the metric perturbations, we can neglect the term stemmed from gauge interactions in the left hand side of \eqref{eq: metrictensor} since ~$\phi$ ~is sufficiently smaller than Planck scale.
 In fact this is a good aproximation as we show the dynamics of metric perturbations later.
 Thus, the metric mode function is obtained as
\begin{equation}
\psi^\pm_{\bm{k}}(x) \approx \psi^\pm_{\text{vac}\bm{k}}(x) + 2\int_0^{\infty}dx'G_\text{vac}(x, x')\left( \dfrac{\phi}{x'}\partial_{x'} + m_\phi(m_\phi \pm x')\dfrac{\phi}{x'^2} \right) t^\pm_{\bm{k}}(x') \label{eq: metmet} \ ,
\end{equation}
where $\psi^\pm_{\text{vac}\bm{k}}(x)$ is a vacuum mode fluctuation in de Sitter space,
\begin{equation}
\dfrac{d^2\psi^\pm_{\text{vac}\bm{k}}}{dx^2}+\left( 1-\dfrac{2}{x^2} \right)\psi^\pm_{\text{vac}\bm{k}} = 0 \label{eq: VAC} \ ,
\end{equation}
and ~$G_\text{vac}(x, x')$ ~is its Green function,
\begin{equation}
\left(\dfrac{d^2}{dx^2} + 1-\dfrac{2}{x^2} \right)G_\text{vac}(x, x') = -\delta(x-x') \ .
\end{equation}
 Thus, we can see that the contribution of gauge fields as well as vacuum metric fluctuations can produce gravitational waves.
 Hence, we need to know the dynamics of the gauge field perturbations.
 For simplicity, we ignore the backreaction of metric perturbations 
\begin{equation}
\dfrac{d^2 t^\pm_{\bm{k}}}{dx^2} + \left(1+\dfrac{m}{x^2} \pm \dfrac{m_t}{x} \right)t^\pm_{\bm{k}} = 0 \label{eq: free} \ .
\end{equation}
\begin{figure}[h]
\begin{center}
\includegraphics[width=12cm,height=12cm,keepaspectratio]{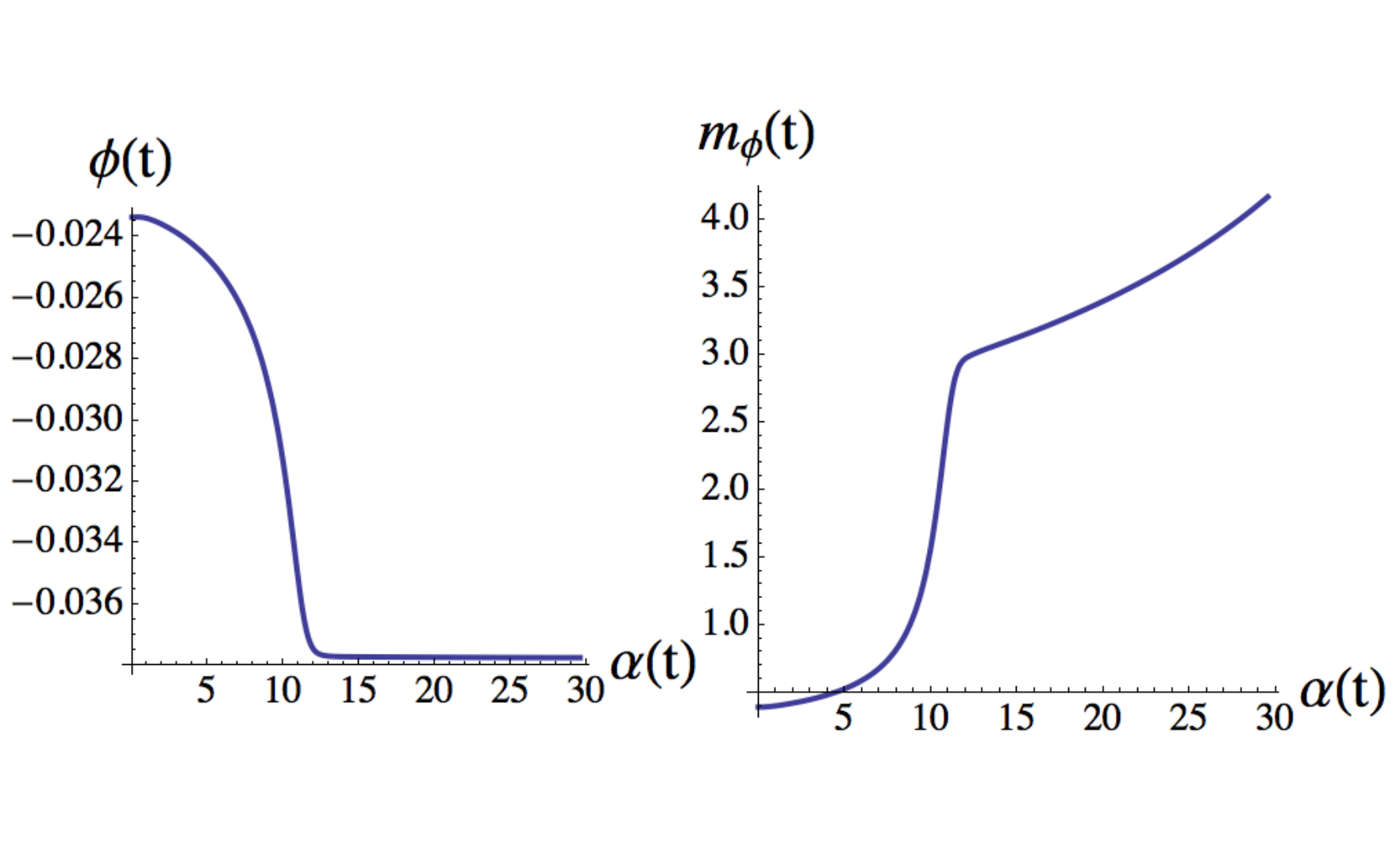}
\end{center}
\caption{The time evolution of the gauge field $\phi(t)$ and $m_\phi(t)$ with ~$(\tilde{\chi}_i \ , \tilde{\omega}_i) = (\pi/2 \ , \pi/9)$ for the sets of parameters $( \ f, \ h, \ \mu, \ g, \ \lambda_\omega \ ) = ( \ 3.5, \ 3.5\times10^{-4}, 10^{-2}, \ 10^{-3}, \ 1.5\times10^{3} \ )$.
 Note that $\alpha(t)$ represents e-folds.
 We see $m_\phi$ is small on CMB scales.
 Consequently, production of chiral gravitational waves is suppressed. 
Interestingly, $m_\phi$ increases during chromo-natural inflation.}
\label{fig: phim}
\end{figure}%
 From this free equation, we can see that ~$t^-_{\bm{k}}$ ~has a tachyonic mass in the following time interval
\begin{equation}
\dfrac{1}{2}(m_t - \sqrt{m_t^2 - 4m}) < x < \dfrac{1}{2}(m_t + \sqrt{m_t^2 - 4m}) \label{eq: period} \ .
\end{equation}
 Since this instability occurs near the horizon-crossing and has a sufficiently large growth rate, we can safely neglect the backreaction of metric perturbations.
 It is easy to see that the interval \eqref{eq: period} depends on $m_\phi$ and gets minimum value when ~$m_\phi \approx 0.8$.
On the other hand, ~$t^+_{\bm{k}}$~ has no instability and its affect to the metric perturbation is negligible as we show later. 
 Thus, we can expect that the dynamics of $m_\phi$ determines the growth rate of one helicity mode of gauge fields, namely, the amount of chiral gravitational waves.

\begin{figure}[h]
\begin{center}
\includegraphics[width=14cm,height=9cm,keepaspectratio]{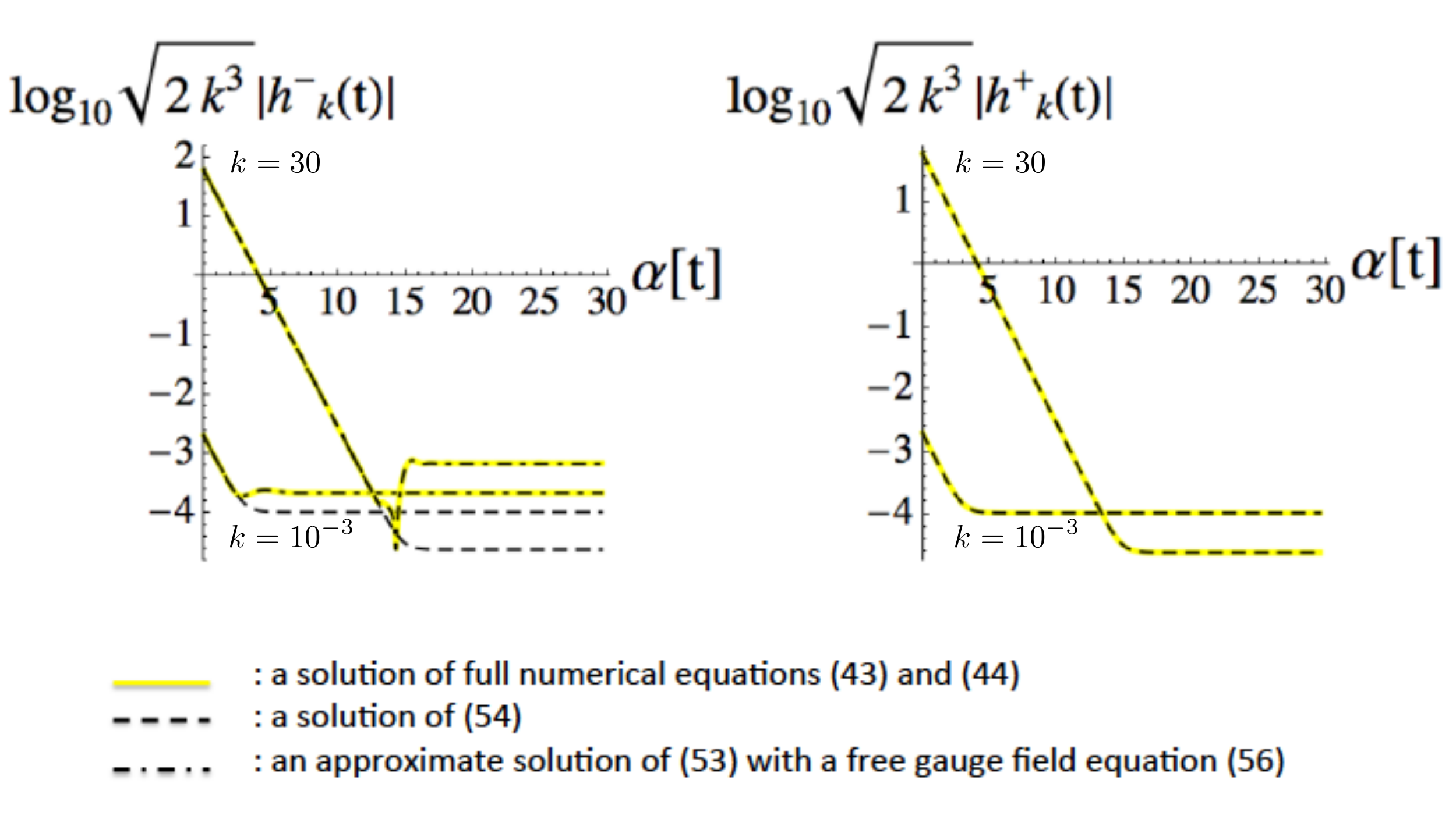}
\end{center}
\caption{ We plotted the time evolutions of the amplitude of physical metric fluctuations $h^{\pm}_{\bm{k}}(t) = \psi^{\pm}_{\bm{k}}(t)/a$ with ~$(\tilde{\chi}_i \ , \tilde{\omega}_i) = (\pi/2 \ , \pi/9)$ for a set of parameters $( \ f, \ h, \ \mu, \ g, \ \lambda_\omega \ ) = ( \ 3.5, \ 3.5\times10^{-4}, 10^{-2}, \ 10^{-3}, \ 1.5\times10^{3} \ )$.
 The left figure is a time evolution of ~$h^-_{\bm{k}}(t)$~ and the right figure is that of ~$h^+_{\bm{k}}(t)$~.
 Here, $\alpha(t)$ is the number of e-foldings.
 We compare the evolutions of fluctuations with different wave-numbers ($k=10^{-3},  30$).
 The fluctuations crossing the horizon before $\alpha(t) \sim 10$ is on CMB scales where natural inflation is dominant.
 On the other hand, the fluctuations crossing the horizon after $\alpha(t) \sim 10$ is on smaller scales where chromo-natural inflation becomes to dominate (see FIG.\ref{fig: phim}). 
 Solid yellow lines are plots of metric fluctuations ~$h^\pm_{\bm{k}}(t)$~ which are solutions of full numerical equations \eqref{eq: tenmetric} and \eqref{eq: tengauge}, while dashed black lines are plots of metric fluctuations ~$h^\pm_{\bm{k}}(t)$~ which are solutions of \eqref{eq: VAC} : standard vacuum metric fluctuations with no contribution of gauge fields.
 Dotdashed black lines in the left figure are plots of ~$h^-_{\bm{k}}(t)$~ which is an approximate solution of \eqref{eq: metmet} with ~$t^\pm_{\bm{k}}(x)$~ satisfying a free field equation \eqref{eq: free}, so we can see that these solutions are good approximation to the full solutions of metric perturbations.
 As initial conditions, we imposed the Bunch-Davies vacuum conditions.
 On CMB scales, ~$h^{-}_{\bm{k}}(t)$~ is close to that with no gauge contribution.
 On the other hand, on smaller scales, $h^{-}_{\bm{k}}(t)$ gets enhanced. }
\label{fig: metricpert}
\end{figure}%

 The time evolutions of $\phi$ and $m_\phi$ are shown in FIG.\ref{fig: phim} , in which $\alpha(t)$ is the number of e-foldings.
 Indeed, we found that $m_\phi$ stays near the minimum $m_\phi \sim 0.8$ 
 on CMB scales, so that we can avoid the overproduction of chiral gravitational waves.
 One may worry the stability of scalar fluctuations in chromo-natural inflation \cite{Adshead:2013qp, Adshead:2013nka}.
 However, natural inflation occurs for the first tens of e-foldings in typical cases, hence we would not need to worry about the instability.
 Remarkably, we can see that  $m_\phi$ gets large value after the end of natural inflation because Hubble parameter gets smaller.
 Therefore, sizable chiral primordial gravitational waves will be produced, which
implies the possibility of detecting chiral gravitational waves on small scales.
 We can see these features in FIG.\ref{fig: metricpert}.
 In FIG.\ref{fig: metricpert}, we plot the time evolution of the amplitude of metric fluctuations at horizon crossing in natural phase and chromo-natural phase, respectively.
 Note that $\alpha(t)$ represents e-folds.
 We can see that the amplitude of $h^{+}_{\bm{k}}(t)$ is nearly the same as the vacuum fluctuation with no contribution of gauge fields, while the amplitude of $h^{-}_{\bm{k}}(t)$ starts to appear in chromo-natural stage due to the chiral enhancement of gauge fields ($ \alpha(t) \sim 15 $).
 Moreover, we verify that the solution of a metric fluctuation $h^{-}_{\bm{k}}(t)  = \psi^{-}_{\bm{k}}(t)/a$ satisfying an equation \eqref{eq: metmet}, where gauge fluctuations $t^\pm_{\bm{k}}(x)$ satisfy a free equation \eqref{eq: free}, is an excellent approximation to the solution of the full equations \eqref{eq: tenmetric} \eqref{eq: tengauge}.

 Finally, we roughly estimate the density parameter of gravitational waves $\Omega_{\text{gw}}(f)$.
 In the case of FIG. \ref{fig: metricpert}, the chiral primordial gravitational waves are produced in the frequency range higher than $10^{-11}$ Hz.
 In the conventional quasi-de Sitter inflation, $\Omega_{\text{con}}(f)$ is given by \cite{Maggiore:1999vm}:
\begin{equation}
h_0^2\Omega_{\text{con}}(f) \approx 10^{-13}\left( \dfrac{H}{10^{-4}} \right)^2 \ ,
\end{equation}
where $h_0 \sim 0.7$ is the dimensionless Hubble parameter.
 In our model, the Hubble parameter $H$ is approximately $10^{-5}$ on small scales.
 Hence,  $h_0^2\Omega_{\text{vac}}$ is about $10^{-15}$.
  On the other hand, the amplitude of chiral gravitational wave is enhanced by the factor $10^{1.5}$ compared to the conventional 
models ( see FIG.3). Since
 $\Omega_{\text{gw}}(f)$ is proportional to the square of amplitude of fluctuations, it can be estimated as $10^{-15}\cdot (10^{1.5})^2 = 10^{-12}$.
 Thus, this gives rise to the density parameter $h_0^2\Omega_{\text{gw}}(10^{-11}~\text{Hz})\sim10^{-12}$.
 Moreover, the density parameter $\Omega_{\text{gw}}$ becomes more large in the high frequency region from mHz to kHz, which can be observed by
 various detectors~\cite{Moore:2014lga} such as
 the space interferometer detector DECIGO \cite{Seto:2001qf} ($h_0^2\Omega_{\text{gw}} \gtrsim 10^{-20}$),  eLISA \cite{AmaroSeoane:2012km} ($h_0^2\Omega_{\text{gw}} \gtrsim 10^{-10}$),
  advanced LIGO,  KAGRA ($h_0^2\Omega_{\text{gw}} \gtrsim 10^{-4}$), and possibly even by the pulsar timing array SKA \cite{Carilli:2004nx} ($h_0^2\Omega_{\text{gw}} \gtrsim 10^{-14}$). It should be stressed that there is a chance to observe chiral primordial gravitational waves~\cite{peloso}.

\section{Conclusion}

 We studied chromo-natural inflation in the axiverse and discussed generation of chiral primordial gravitational waves.
 Concretely speaking, we investigated natural inflation with two axions coupled with a SU(2) gauge field.
 We assumed hierarchy between the axion coupling constants to gauge fields. Note that the coupling constants need not be small,
rather one of them should be large so that chromo-natural inflation occurs. 
 We assumed the axion which has strong coupling constant to the gauge field is energetically sub-dominant initially.
 This is natural because there are many axions except for the one with strongly coupled axion.
 Then, we found that conventional natural inflation commences, continues for tens of e-foldings, and
 subsequently, chromo-natural inflation takes over from natural inflation.
 Since the role of gauge fields in the early stage is negligible,  we expect that the predictions are in agreement with observations on CMB scales.
 Thus, we concluded that overproduction of chiral gravitational waves in chromo-natural inflation model can be circumvented in the axiverse.
 This is because the parameter of the gauge field $m_\phi$ is sufficiently small during natural inflation.
 We also found that $m_\phi$ increases in the course of chromo-natural inflation, which implies that chiral gravitational waves are enhanced on small scales. Remarkably, the chiral primordial gravitational waves are produced 
 in an interesting frequency range higher than $10^{-11}$Hz, which might be detectable in future gravitational wave observations such as
 pulsar timing experiment using SKA, advanced LIGO, eLISA, and KAGRA. Although other mechanisms can also produce chiral primordial gravitational waves~\cite{Lue:1998mq,Satoh:2007gn}, since the spectrum of primordial gravitational waves in chromo-natural inflation is distinctive,
the observations at several different frequencies will be able to discriminate the chromo-natural inflation with two axions from others. 

Note that we treated natural inflation with super-Planckian decay constant.
However, by using the mechanism in \cite{Kim:2004rp, Kappl:2014lra} where a combination of two axions is used,
 we are able to get an ``effective'' super-Planckian decay constant from two sub-Planckian decay constants.
Hence, in the context of multi-field extension of natural inflation, our model is phenomenologically viable.

 For future work, we need to analyze the spectrum of chiral gravitational waves in detail to compare the predictions with observations.
 We should also study scalar perturbations and explicitly check stability  of natural inflation with two axions~\cite{Adshead:2013qp, Adshead:2013nka}.
 Moreover, we will seek a way to embed the current model into fundamental theory.
 We leave these issues for future work.

\section*{Acknowledgments}
We would like to thank Kei Yamamoto for fruitful discussions.
This work was supported by  Grants-in-Aid for Scientific Research (C) No.25400251
 and Grants-in-Aid for Scientific Research on Innovative Areas No.26104708.

\end{document}